\documentclass[11pt,twocolumn]{emulateapj_Astroph}
\usepackage{color}
\usepackage{ulem} % AV
\usepackage{epsfig,amsmath,natbib}
\usepackage{epstopdf}

\newcount\listnorom
\newcommand\listromanDE{\global\advance \listnorom by 1
{\lowercase\expandafter{(\romannumeral\listnorom)}\ }}

\newcount\listnumber
\newcommand\listDE{\global\advance \listnumber by 1
{\lowercase\expandafter{(\number\listnumber)}\ }}
\newcommand\newlistDE{\listnumber=0}
\newcommand{\Rgp}{R_\mathrm{gp}}
\newcommand{\Omir}{\omega_\mathrm{mir}}
\newcommand{\Amir}{A_\mathrm{mir}}
\newcommand{\FcrUp}{f^\mathrm{ep}}
\newcommand{\FcrIso}{f^\mathrm{ep}_\mathrm{iso}}
\newcommand{\gamray}{$\gamma$-ray}

\newcommand{\Xone}{\xi_1}

\newcommand{\FpfT}{f_\mathrm{pf}^\mathrm{ep}(z,p,\mu)}
\newcommand{\vpf}{v_\mathrm{pf}}
\newcommand{\Nscat}{N_\mathrm{scat}}
\newcommand{\PAS}{pitch-angle-scattering}
\newcommand{\pas}{pitch-angle-scattering}
\newcommand{\msd}{mean-square-displacement}
\newcommand{\aD}{A_D}
\newcommand{\SupD}{Super-diffusion}
\newcommand{\supD}{super-diffusion}
\newcommand{\supDif}{super-diffusive}
\newcommand{\SupDif}{Super-diffusive}

\newcommand{\zLF}{z_\mathrm{LF}}
\newcommand{\LLF}{\lambda_\mathrm{LF}}

\newcommand{\Lz}{\lambda_0}
\newcommand{\mhd}{magnetohydrodynamic}

\newcommand{\rgz}{r_{g0}}

\newcommand{\jCR}{\mathbf{j}^\mathrm{ep}}
\newcommand{\nCR}{n_\mathrm{ep}}
\newcommand{\SC}{self-consistent}
\newcommand{\SCly}{self-consistently}

\newcommand{\EP}{energetic particle}

\newcommand{\nonrel}{nonrelativistic}
\newcommand{\rel}{relativistic}
\newcommand{\levy}{{\it L\'evy}}

\newcommand{\mc}{Monte Carlo}
\newcommand{\MC}{Monte Carlo}
\newcommand{\Facc}{Fermi acceleration}
\newcommand{\DSA}{diffusive shock acceleration}

\newcommand{\MFA}{magnetic field amplification}
\newcommand{\alf}{Alfv\'en}

\newcommand{\TP}{test-particle}
\newcommand{\NL}{nonlinear}
\newcommand{\Fcr}{F_\mathrm{ep}}
\renewcommand{\baselinestretch}{1.5}

\newcommand{\valf}{v_\mathrm{alf}}

\newcommand{\Lfeb}{z_\mathrm{FEB}}
\newcommand{\pcc}{cm$^{-3}$}
\newcommand{\kmps}{km s$^{-1}$}

\newcommand{\muG}{$\mu$G}

\newcommand{\SA}{semi-analytic}
\newcommand{\FEB}{free escape boundary}
\newcommand{\RH}{Rankine-Hugoniot}

\newcommand{\mfp}{mean free path}

\newcommand{\Bls}{B_\mathrm{ls}}

\newcommand{\rg}{r_g}
\newcommand{\rgZ}{r_{g0}}

\newcommand{\delMax}{\delta\theta_\mathrm{max}}
\newcommand{\delAng}{\delta\theta}
\newcommand{\deltime}{\delta t}

\newcommand{\Rtot}{R_\mathrm{tot}}
\newcommand{\Rsub}{R_\mathrm{sub}}

\newcommand{\Vscat}{v_\mathrm{scat}}
\newcommand{\vscat}{v_\mathrm{scat}}

\newcommand{\Fth}{F_\mathrm{th}}

\newcommand{\Fw}{F_w}
\newcommand{\Pth}{P_\mathrm{th}}
\newcommand{\Pcr}{P_\mathrm{ep}}
\newcommand{\Pw}{P_w}

\newcommand{\PhiP}{\Phi^\mathrm{part}_P}
\newcommand{\PhiPz}{\Phi_{P0}}
\newcommand{\PhiE}{\Phi^\mathrm{part}_E}
\newcommand{\PhiEz}{\Phi_{E0}}
\newcommand{\Qesc}{Q_\mathrm{esc}}
\newcommand{\qesc}{q_\mathrm{esc}}

\newcommand{\kmax}{k_\mathrm{max}}
\newcommand{\kmin}{k_\mathrm{min}}
\def\lsim{\;\raise0.3ex\hbox{$<$\kern-0.75em\raise-1.1ex\hbox{$\sim$}}\;}
\def\gsim{\;\raise0.3ex\hbox{$>$\kern-0.75em\raise-1.1ex\hbox{$\sim$}}\;}
\def\aap{A\&A}                   % "Astron. Astrophys."
\def\apjs{ApJS}                  % "Astrophys. J. Suppl. Ser."
\def\apjl{ApJ}                   % letter at ApJ

\def\lsim{\;\raise0.3ex\hbox{$<$\kern-0.75em\raise-1.1ex\hbox{$\sim$}}\;}
\def\gsim{\;\raise0.3ex\hbox{$>$\kern-0.75em\raise-1.1ex\hbox{$\sim$}}\;}
\def\aap{A\&A}

\def\apjl{ApJ Letters}

\newcount\Inum
\newcount\IInum
\newcount\IIInum
\newcount\IVnum
\def\I{\global\multiply\IInum by 0 \global\multiply\IIInum by 0
            \global\multiply\IVnum by 0 \global\advance \Inum by 1
            {\the\Inum. }}
\def\II{\global\multiply\IIInum by 0\global\multiply\IVnum by 0
       \global\advance \IInum by 1 {\the\Inum.\the\IInum. }}
\def\III{\global\multiply\IVnum by 0\global\advance \IIInum by 1
            {\the\Inum.\the\IInum.\the\IIInum. }}
\def\IV{\global\advance \IVnum by 1
            {\the\IVnum. }}
\shorttitle{Super-diffusion in nonlinear shock acceleration}
\shortauthors{Bykov, Ellison \& Osipov}

\begin{document}

\title{A nonlinear  Monte Carlo model of super-diffusive shock acceleration \\ with magnetic field amplification \vskip12pt}
\author{Andrei M. Bykov,\altaffilmark{1,2,3}
Donald C. Ellison\altaffilmark{4} and
Sergei M. Osipov\altaffilmark{1}
}
\altaffiltext{1}{Ioffe Institute for Physics and Technology, 194021
St. Petersburg, Russia; ambykov1@gmail.com}

\altaffiltext{2}{Saint-Petersburg State Polytechnical University,
Saint-Petersburg, Russia}

\altaffiltext{3}{International Space Science Institute, Bern, Switzerland}

\altaffiltext{4}{Physics Department, North Carolina State
University, Box 8202, Raleigh, NC 27695, U.S.A.;
don\_ellison@ncsu.edu}

%\author{Andrei M. Bykov}
%\affiliation{Ioffe Institute, St. Petersburg
%194021,  St.Petersburg State Polytechnical University, Russia,
%International Space Science Institute, Bern, Switzerland;
%byk@astro.ioffe.ru}
%\author{Donald C. Ellison }
%\affiliation{North Carolina State University, Department of Physics,
%Raleigh, NC 27695-8202, USA; ellison@ncsu.edu}
%\author{Sergei M. Osipov}
%\affiliation{Ioffe Institute, St.Petersburg 194021}

\begin{abstract}
Fast collisionless shocks  in cosmic plasmas  convert their kinetic energy flow into the hot downstream
thermal plasma with a substantial fraction of energy going into a broad spectrum of superthermal charged particles and  magnetic fluctuations.
The superthermal particles can penetrate into the shock upstream
region producing an extended shock precursor. The cold upstream
plasma flow is decelerated by the force provided by the superthermal
particle pressure gradient. In high Mach number collisionless
shocks, efficient  particle acceleration is likely coupled with
turbulent magnetic field amplification (MFA)  generated by the
anisotropic distribution of accelerated particles.
This anisotropy is determined by the fast particle transport making the problem strongly nonlinear and
multi-scale.
Here, we present a nonlinear  Monte Carlo model of collisionless
shock structure with super-diffusive propagation of
high-energy Fermi accelerated particles  coupled to particle acceleration and MFA which affords a consistent description of strong shocks.
A distinctive feature of  the \mc\ technique is that it includes the
full angular anisotropy of the particle distribution at all
precursor positions. The model reveals that  the super-diffusive transport of energetic particles (i.e., \levy-walk
propagation)  generates a strong   quadruple anisotropy in the precursor
particle distribution.
The resultant pressure anisotropy of the high-energy particles produces a non-resonant
mirror-type instability which amplifies compressible wave modes with
wavelengths longer than the gyroradii of the highest energy protons
produced by the shock.
\end{abstract}
\pacs{PACS:}
\pacs{Valid PACS appear here}% PACS, the Physics and Astronomy
                             % Classification Scheme.
%\keywords{Neutrino, supernova remnants, stellar winds, interstellar %medium}%Use showkeys class option if keyword
                              %display desired
%%\maketitle

\section{Introduction}
In contrast to collision-dominated shocks, strong collisionless
plasma shocks are capable of converting the kinetic power of the
upstream flow to both thermal and non-thermal components.
The conversion process, in diffuse plasmas where Coulomb collisions
are very infrequent, is due to highly nonlinear interactions between
the particles and the background magnetic turbulence utilizing the
first-order Fermi mechanism
\citep[][]{baloghtreumann13,marcowithea16}.
While collisionless shocks are difficult to study  in the
laboratory, they are known to exist in cosmic plasmas and play  a
critical role in producing nonthermal particles observed throughout
the cosmos.
The lack of Coulomb collisions allows the  nonthermal components --
accelerated energetic particles {(EP)} and magnetic turbulence -- to
be long-lived  and dynamically significant. 
Observations and theory
both confirm that  particle acceleration can be efficient enough so
the energetic particles that  penetrate into the shock precursor can
slow the bulk supersonic flow significantly before the viscous
subshock occurs. 
The viscous subshock is a small-length-scale collisionless shock of moderate Mach number $\sim 3$. The subshock, which is directly observed in heliospheric shocks and particle-in-cell (PIC) simulations, involves mainly thermal particles and is required to produce the entropy and jumps in plasma density, temperature, and magnetic field needed to satisfy the \RH\ relations.
The superthermal precursor particles are highly anisotropic and drive
a number of plasma instabilities
\citep[e.g.,][]{bell04,schureea12,bbmo13}
producing magnetic turbulence which interacts with the precursor  particles producing the turbulence.
Modeling strong collisionless shocks  is an interesting multi-scale
problem strictly from the  point of view of nonlinear plasma
physics. It is also fundamentally important for understanding
energetic particle populations observed in space.

The fact that collisionless shocks accelerate  ambient particles in
many locations ranging from the Earth bow shock to shocks in
galactic clusters is widely accepted. While a great deal is known
about the acceleration process, the most important part---the
collisionless wave-particle interactions driving particle
isotropization---remains uncertain. In principle, a full description
of the plasma interactions is obtainable with PIC 
simulations \citep[e.g.,][]{PCS2015}.
However, PIC simulations are computationally  expensive and results
thus far are limited to a relatively narrow dynamical range which is
particularly restricting for the modeling of \nonrel\  shocks such
as those seen in supernova remnants.

All collisionless shock calculations, other than PIC,
 must approximate particle transport
and most models assume fast particles obey standard diffusion (in
the local plasma rest frame) where the \msd\ is proportional to
time, i.e.,
\begin{equation} \label{eq:diff}
\left < \Delta z^2 \right > = \aD t^b
\ ,
\end{equation}
with $b=1$. This simple equation (even with $b=1$) hides  a great
deal of complexity since the proportionality factor, $\aD$, depends
non-linearly on the details of the self-generated magnetic
turbulence and will vary with particle momentum, position relative
to the subshock, and the \Facc\ efficiency.

Furthermore, there is no fundamental reason why $b=1$ in
Eq.~(\ref{eq:diff}) and both super- ($b>1$) and sub-diffusive
($b<1$) transport regimes are possible in complex nonlinear and
intermittent systems
\citep[e.g.,][]{Levy_Nature93,Levy_intermitt,ZM04,LevyRMP15}.
Such non-standard diffusion has been shown to be important in
 laboratory  and fusion plasmas
\citep[e.g.,][]{Balescu2005,rev_nonclas_transp13,
nonclass_transp_tokamak15}.
Furthermore, there is evidence  from in-situ spacecraft observations of
heliospheric shocks for anomalous diffusion where the \msd\ grows
non-linearly with time with $b>1$ \citep[][]{perriea15}.
The effect of complex transport on EP propagation and acceleration
was discussed by
\citep[][]{kirk_ea_AA96,peri_zimbardo09,malkov_diam09,ZP13,lazarian_yan14}.

We consider Fermi acceleration in strong quasi-parallel shocks where the average magnetic field direction upstream of the shock is close to the shock normal, implicitly assuming the magnetic field at the subshock is turbulent enough so effects from perpendicular components of the field can be ignored \citep[e.g.,][]{EBJ95}. 
An analysis of obliquity effects on particle propagation within the \mc\ model suggests  
that shocks can be considered ``parallel" for angles  up to $\sim \pi/4$ from the shock normal.
The shock  produces anisotropic EP distributions in the
shock precursor that result in strong non-adiabatic amplification
of irregular magnetic fields by EP-driven instabilities
\citep[e.g.,][]{marcowithea16}.
Magnetic field fluctuations present in the interstellar medium are highly amplified by these
instabilities as they traverse the shock precursor.

An important characteristic of strong shocks undergoing efficient Fermi acceleration is that the highest energy particles are distributed throughout the entire
precursor while lower energy particles are
concentrated close to the subshock.
This results in a strong spatial
dependence of the  growth rate of magnetic fluctuations with a given
wavenumber and may lead to \supD\ in the outer precursor where the
magnetic turbulence is growing and is likely highly intermittent.

Here, we assume the  intermittency of the  turbulence dominates the
EP propagation in the \supD\ region of the precursor before the
growing turbulence saturates.
The boundary of the \supD\ region in the
upstream is parameterized by $\zLF$ which is the distance from the
subshock.
This parameterization is needed since we use simplified
models for magnetic turbulence cascade, which deal only with the spectrum of turbulence, while the intermittency requires a more
detailed description. This is discussed in more detail in
Section~\ref{sec:model}.

\SupDif\ EP propagation has been seen at a few gyro-rotation periods
in \mhd\ (MHD) simulations of the non-resonant EP-driven
 instability in \citep{reville08} where the EP current was fixed
 (i.e., without considering \SC\ EP evolution).
However, the question whether EP transport is diffusive in the shock
precursor when EP-driven instabilities are rapidly growing has not
yet been addressed.
The \NL\ backreaction of EPs on the shock structure will
influence the EP transport and anisotropy making \NL\ calculations
coupling EP production, shock structure, instability growth, and
particle transport essential.
While in principle, the full time-dependent picture can be
modeled with large-scale PIC simulations, such calculations are still beyond current capabilities.
Until these calculations become available, the effect of
non-diffusive propagation on \NL\ Fermi shock acceleration can be studied with Monte Carlo
simulations of particle transport
by simply assuming that the EP transport in some regions
of the shock precursor is non-diffusive.

An important distinction with Fermi shock acceleration models which are based on
the advection-diffusion transport  equation is that they
model only EP-current
anisotropies. The \mc\ model does not make a diffusion approximation
and thus can account for arbitrary angular anisotropy
harmonics.  We find that the inclusion of \supD\ from \levy-walk
scattering-length distributions produces specific anisotropies
in the particle distributions that drive instabilities that do not occur with standard diffusion models.

When Fermi shock acceleration is efficient, particle transport, the shock  structure,
MFA, turbulence cascading, and thermal particle injection  must all
be calculated \SCly. No technique, not even PIC simulations, can
currently do this full calculation from first principles over a
dynamic range sufficient to model EP production in SNRs or other
strong \nonrel\ shocks.
While approximations must be made, much of the essential \NL\ physics can be modeled with \mc\ techniques.

The \mc\ simulation we use couples analytic  descriptions
of resonant and non-resonant wave growth with anisotropic particle
transport in EP-dominated shocks
\citep[see][and references therein]{EBJ96,beov14}.
All of the \NL\ effects mentioned above have been consistently
included assuming standard diffusion. We  now generalize the \mc\
model by explicitly including \supD\ in the shock precursor.
Because the \mc\ model accounts for the full anisotropic EP distribution
functions, the dispersion relations we derive simultaneously include the EP-driven resonant streaming instability, and the two EP-current driven
instabilities: Bell's short-wavelength instability, and the long-wavelength instability
(see \citep{schureea12,bbmo13}).

Our \NL\ model shows two distinctive features.
The first is that  \supD\ results in  a highly amplified specific
quadrupole anisotropy of EP particles.
This anisotropy produces a
mirror instability that has not been previously considered in
efficient shock acceleration. The mirror instability  contributes significantly to the generation of long-wavelength magnetic turbulence which, in principle, can be studied with direct measurements of heliospheric shocks and with
indirect analysis of the broadband synchrotron emission seen in supernova remnants.
The  second feature is that \supD\ in the shock  precursor results in a
substantial broadening of the spectrum of energetic particles escaping the
precursor. Energetic particles escaping the shock will undergo inelastic collisions with surrounding matter and produce high-energy radiation.
We quantify the mirror instability and escaping EPs with a limited number of \mc\ examples.

\newlistDE

\section{The nonlinear Monte Carlo shock model} \label{sec:model}
We construct a steady-state model of a plane-parallel,   \nonrel\
collisionless  shock where the \NL\ shock structure is determined
iteratively. The shock is directed in the $-z$-direction with a
subshock at $z=0$ and an upstream \FEB\ (FEB) limiting \Facc\ at $z
= \Lfeb$.
For simplicity, we model only protons since they mainly determine the \NL\ shock structure and drive the
long-wavelength instabilities we consider. Electrons can be included, as in \cite{WarrenEllison2015}, when radiation is calculated.

The \mc\ shock model includes the following main elements: \\
\listDE Particle injection, which is \SCly\ coupled to \Facc\ where some fraction of
shock-heated thermal particles re-cross the subshock (assumed transparent), gain additional energy, and enter the acceleration process. Any particle that crosses from downstream back upstream at least once is termed energetic; \\
\listDE Shock-smoothing, where backpressure from  superthermal particles slows and heats the precursor plasma upstream of the viscous subshock in order to conserve momentum and energy; \\
\listDE The \SC\  determination of the overall shock compression ratio, $\Rtot$, taking into account escaping EPs, magnetic pressure, and the modification of the equation of state from the production of \rel\ particles; \\
\listDE Fluctuating magnetic fields simultaneously  calculated from resonant, short-wavelength, long-wavelength, and mirror instabilities generated from the EP current, and \supD\ pressure anisotropies in the shock precursor; \\
\listDE Momentum and position dependent particle transport determined from the self-generated magnetic turbulence; \\
\listDE A determination of the local plasma scattering center speed relative to the bulk plasma  from energy conservation without assuming \alf\ waves; and, \\
\listDE Turbulence convection and compression, cascade, and
dissipation of wave energy into the background plasma.
The iterative \mc\ approach allows all of these processes to be coupled and calculated simultaneously in a reasonably
consistent fashion.

\section{Mass-Energy-Momentum Conservation} \label{sec:III}
We determine the \SC\ shock structure with an iterative procedure by
forcing mass-energy-momentum conservation. All particles--thermal
and superthermal--are transported through the shock keeping full
account of the anisotropic particle distribution and the momentum
and energy contributions from the magnetic fluctuations (see
\cite{beov14} for full details).

In the shock rest frame, the mass flow conservation is given by
\begin{equation}\label{massFlux}
\rho(z)u(z)=\rho_{0}u_{0} \ ,
\end{equation}
where  $\rho(z)$ is the plasma density, $u(z)$ is the bulk flow speed, and the subscript ``0" here
and elsewhere indicates far upstream values. We define the ``shock structure" as $u(z)$, where $z$ is the distance measured from the subshock at $z=0$.
The momentum flux conservation is determined by
\begin{equation}\label{momentunFlux}
\PhiP(z)+P_{w}(z)=\PhiPz \ ,
\end{equation}
where $\PhiP(z)$ is the particle momentum flux, $P_{w}(z)$ is the
momentum flux carried by the magnetic turbulence, and $\PhiPz$ is the far
upstream momentum flux, i.e., upstream from the free escape boundary
where the interstellar magnetic field is $B_0$.

Separating the contributions from the thermal and accelerated
particles we have
\begin{equation}\label{momentunFlux1}
\rho(z)u^{2}(z) + \Pth(z) + \Pcr(z) + \Pw(z)=\PhiPz \ ,
\end{equation}
where $\Pth(z)$ is the thermal particle pressure and $\Pcr(z)$ is
the accelerated particle pressure. As mentioned above, a particle
is ``accelerated" if it has crossed the subshock more then once.
There is no other injection threshold and even though we use the
subscript ``EP", the vast majority of accelerated particles will
always be \nonrel.
Of course, if the acceleration is efficient, a large fraction of the
pressure may be in \rel\ particles.

The energy flux conservation law is
\begin{equation}\label{momentunEnergy}
\PhiE(z) + F_{w}(z)=\Phi_{E0} \ ,
\end{equation}
where $\PhiE(z)$ and  $F_{w}(z)$ are the energy fluxes in particles
and magnetic field correspondingly, and $\PhiEz$  is the energy flux
far upstream.
Taking into account particle escape at an upstream FEB, this can be
re-written as
\begin{equation} \label{momentunEnergy1}
\frac{\rho(z)u^{3}(z)}{2} + \Fth(z) + \Fcr(z) + \Fw(z) + \Qesc =
\PhiEz \ ,
\end{equation}
where $\Fth(z)$ is the internal energy flux of the background
plasma, $\Fcr(z)$ is the energy flux of accelerated particles, and
$\Qesc$ is the energy flux of particles that escape at the upstream
FEB (note that $\Qesc$ is defined as positive even though EPs escape
moving in the negative $z$-direction).

The separation between ``thermal" particles and ``accelerated" particles
in a shock undergoing \DSA\ is not necessarily well defined.
Furthermore, the energy exchange between the thermal and
superthermal populations is certain to occur through non-trivial
wave-particle interactions.
Nevertheless, the bulk of the plasma mass will always be in
quasi-thermal background particles and the internal energy flux of
this background plasma can be expressed as
\begin{equation}\label{F_th_P_th}
\Fth(z) = u(z)\frac{\gamma_{g} \Pth(z)}{\gamma_{g} -1} \ ,
\end{equation}
where $\gamma_{g}=5/3$ is the adiabatic index of the background thermal
plasma.

All of the quantities in Eqs.~(\ref{massFlux})--(\ref{F_th_P_th}) are directly measured in the \mc\ simulation. The magnetic turbulence, and therefore $\Pw(z)$ and $F_{w}(z)$,  is determined from $u(z)$ and the analytic expressions for wave growth and cascading discussed below. Once the assumptions for wave growth are made, the equations for mass, momentum, and energy flux are over determined and a unique, \NL\ solution conserving mass, momentum, and energy flux in the shock rest frame can be found by iterating $u(z)$. In practice, a ``consistent solution" is accepted when the momentum and energy fluxes are conserved to within a few percent at all $z$.

\subsection{Turbulence cascade}\label{MTC}
The magnetic turbulence energy flow $\Fw(z)$ in
Eq.~(\ref{momentunEnergy1}) is determined by the spectral energy
density of the magnetic fluctuations $W(z,k)$  
(see \cite{beov14} for details ) 
which obeys the equation
\begin{eqnarray}\label{eqvW}
   & & u(z)\frac{\partial W\left(z,k\right)}{\partial z}
      +\frac{3}{2}W\left( z,k \right)\frac{du(z)}{dz}
      +\frac{\partial\Pi\left(z,k\right)}{\partial k}  =\\
   & & = G(z,k) - \mathcal{L}(z,k) \nonumber,
\end{eqnarray}
where $\Pi(z,k)$ is the flux of magnetic energy, $G(z,k)$ is the
spectral energy growth rate due to EP instabilities, and
$\mathcal{L}(z,k)$ is the turbulence dissipation rate.
Following \cite{Matthaeusea09}, we approximate the turbulent energy
cascade rate as
\begin{equation}\label{flux_Pi}
\Pi(z,k) =
-\frac{D_{\rm K}}{\sqrt{\rho(z)}}k^{\frac{11}{2}}W(z,k)^{\frac{1}{2}}\frac{\partial}{\partial
k}\left[\frac{W(z,k)}{k^{2}}\right]
\ ,
\end{equation}
where $D_{\rm K}=0.14$ is the cascade constant which was chosen
to match the Kolmogorov constant.
To study  the effect of anisotropic turbulent cascade  we simulated
two regimes: one  assumes the  turbulent energy cascade is given by
Eq.~(\ref{flux_Pi}), the other assumes no cascade.
The unperturbed spectrum of turbulence entering the free
escape boundary at $\Lfeb$ is taken to be Kolmogorov, typically assumed for the interstellar medium.  The incoming spectrum is normalized by
\begin{equation}\label{W_FEB_norm}
\int^{\kmax}_{\kmin} W(\Lfeb,k)dk=\frac{B_{0}^{2}}{4\pi}
\ .
\end{equation}

\subsection{Particle propagation model}
With normal (i.e., non-\levy-walk) diffusion the \mc\ simulation
moves particles with a \pas\ scheme that has been described in
\citep{EBJ96}. Briefly, after a time $\deltime$ much less than a gyroperiod a particle scatters
isotropically and elastically in the local plasma frame through a
angle $\delAng \le \delMax$, where $\delAng$ is chosen randomly
between 0 and $\delMax$.
The maximum scattering angle is given by
\begin{equation} \label{eq:Tmax}
\delMax = \sqrt{6 \deltime/t_c}
\ ,
\end{equation}
where  $t_c=\lambda_0/\vpf$ is the collision time,
$\lambda_0(z,p)$ is the position and momentum dependent scattering
length, $\vpf$ is the particle speed in the rest frame of scattering
centers,
$\rg = pc/[e \Bls(z,p)$] is the particle gyro-radius, and
$\Bls(z,p)$ is the local amplified magnetic field
determined by summing fluctuations with wavelengths larger than $\rg$ \citep[see Eq.~(19) in reference][for a full description]{beov14}.
In the normal scattering region we assume Bohm diffusion, i.e., $\lambda_0(z,p)=\rg(z,p)$.

For non-\levy-walk scattering, particles always move a fraction of
$\Lz(z,p)$ in the time interval $\deltime$, where $\Lz(z,p)$ is the
mean free path a EP obtains scattering in the self-generated
magnetic turbulence. We use the same \mc\ model as described in
earlier work (see e.g., Section~2.7 of
\citep{beov14}) to calculate MFA and $\Lz(z,p)$  except we now
include the \supD-induced mirror instability along  with the
resonant and non-resonant instabilities.

\subsection{Super-diffusive EP propagation: 
the L\'evy-walk model}\label{sec:SD}
To model \supDif\ particle propagation we use a \levy-walk model which assigns a random scattering length $\LLF$ to determine the path length. The probability density function of the
$\LLF$ values has a power-law asymptotic form:
\begin{equation} \label{ProbDen}
\Psi(\LLF) \propto \LLF^{-\nu}
\quad\textrm{for}\quad
\LLF(z,p) > \lambda_0(z,p)
\ .
\end{equation}
This density function produces  so-called ``heavy tails" for $\nu
\le 3$, where $\nu=2$ is the well known Cauchy distribution. We
describe below the specific algorithms to generate the random
scattering lengths $\LLF$ with the power-law probability density
functions.
For completeness we discuss the Cauchy distribution in
\S~\ref{hCd} noting that since we only consider $\LLF >0$, $\nu=2$
in Eq.~(\ref{ProbDen}) gives the half-Cauchy distribution. We
further note that while Cauchy distributions are used for
mathematical convenience in \SA\ calculations,  they have an
infinite mean and variance. The flexibility of the \mc\ model (and
the fact that particles always have a speed less than $c$) allows us
to derive a more general expression for the scattering length in
\S~\ref{PL_LF} for $2 < \nu \le 3$. We show examples with $\nu=2.1$
because this value gives a finite mean yet  produces results similar
to well-studied Cauchy ones.

\subsubsection{The half-Cauchy distribution}\label{hCd}
With $\nu=2$ we assume the EP scattering length in the \supDif\ region $\Lfeb < z < \zLF$  is determined by
\begin{equation} \label{mfp_LF}
\lambda_{\rm LF}(z,p) =
\tan \left(\frac{\pi \xi_0}{2} \right)\lambda_0(z,p)\quad\mathrm{for}\quad
z < \zLF
\ .
\end{equation}
Here $\xi_0$  are random numbers  uniformly distributed over the
interval  [0,1), where the brackets
indicate values up to but not including 1.
Values of $\xi_0 \sim 1$ give extremely long
scattering lengths, forcing a modification of Eq.~(\ref{eq:Tmax}),
while $\LLF \sim 0$ for $\xi_0 \sim 0$.

Integrating Eq.~(\ref{mfp_LF}) from $p$ to $\infty$ we find the  cumulative distribution function
\begin{equation}\label{cdf}
F(\LLF) =
\frac{2}{\pi} \arctan \left(\frac{\lambda_{\rm LF}}{\lambda_0}\right)
\ ,
\end{equation}
and the probability density function corresponding to
Eq.~(\ref{cdf}) is
\begin{equation}\label{pdf}
 \Psi(\lambda_{\rm LF}) = \frac{2}{\pi \lambda_0}\left[1 + \left(\frac{\lambda_{\rm LF}}{\lambda_0}\right)^2\right]^{-1}
\ .
\end{equation}

\subsubsection{L\'evy-type power-law distributions
with $2<\nu\le 3$}\label{PL_LF}
For $2 < \nu \leq 3$ the  recipe is somewhat more complicated. The
scattering length in the \levy-walk region of the shock precursor where
$z < \zLF$, is given by
\begin{eqnarray} \label{mfp_levy_fly_mu}
& & \lambda_{\rm LF}=\lambda_0\left[\frac{2 \nu \left(\nu -
2\right)\xi_{1}}{\left(\nu -1\right)^{2}}+ 1\right],~~
 \xi_{1}\leq
D_0, \\
& & \lambda_{\rm LF}=\lambda_0\left [ \frac{2(\nu - 2)}{\nu -
1}\left[\frac{1}{\nu \left(1-\xi_{1}\right)}\right] ^{\frac{1}{\nu
-1}}+1\right ], ~~ \xi_{1}> D_0,
\nonumber
\end{eqnarray}
where $D_0=(\nu-1)/\nu$, and  $\xi_{1}$  are random numbers
uniformly distributed over the interval  [0,1).
The corresponding probability  density function for
$\lambda_{\rm LF}^{\ast} \equiv \lambda_{\rm LF}-\lambda_0$ is
 \begin{eqnarray} \label{probability_MFP}
& & \Psi\left( \lambda_{\rm
LF}^{\ast}\right)=\frac{C_{\nu}}{\lambda_{0}}, ~~ {\rm for}~~
\lambda_{\rm
LF}^{\ast} \leq \, \lambda_{\ast},
 \nonumber \\ & &
\Psi\left( \lambda_{\rm LF}\right)=\frac{C_{\nu}}{\lambda_{0}}
\left(\frac{ \lambda_{\rm
LF}^{\ast}}{\lambda_{\ast}}\right)^{-\nu},~~ {\rm for}~~
\lambda_{\rm LF}^{\ast}>\lambda_{\ast} \ ,
\end{eqnarray}
where
\begin{equation}
\lambda_{\ast} = 2 \lambda_{0} \frac{\nu -2}{\nu -1}
\quad\textrm{and}\quad
C_{\nu}=\frac{(\nu -1)D_0}{2(\nu -2)}
\end{equation}
\citep[see][for a full discussion]{Trotta2015}.

The normalization and mean value are determined by
\begin{equation}
\int_{0}^{\infty}\Psi\left(\lambda\right)d\lambda=1
\end{equation}
and
\begin{equation} \label{Eq:mean}
\int_{0}^{\infty}\lambda\Psi\left(\lambda\right)d\lambda=\lambda_{0}
\ ,
\end{equation}
respectively.
While Eq.~(\ref{mfp_levy_fly_mu}) applies for $2<\nu \le3$, we
restrict  our calculations here to $\nu=2.1$ to ensure
Eq.~(\ref{Eq:mean}) yields a finite mean.

We implement \levy-walk transport in a piecewise continuous way. In
the precursor, downstream from some precursor position $\zLF$,
particles propagate diffusively with a \mfp\ $\LLF(z,p) = \Lz(z,p)$.
Far upstream, between $\zLF$ and the FEB at $\Lfeb$, we assume EPs
propagate super-diffusively. Typically, $\zLF \leq -10^4 \rgz$,
where $\rgz=m_p u_0 c/(e B_0)$ is the gyroradius of a proton with
speed equal to the shock speed $u_0$ in the far upstream magnetic
field $B_0$.
For $z < \zLF$, a particle is given a random scattering length $\LLF(z,p)$
generated with the  \levy-type stable distribution described below.

We restrict  \supDif\ propagation to regions well upstream  from the
subshock where the self-generated turbulence has not reached
saturation levels.  Close to the FEB the turbulence is growing
rapidly from the escaping EP flux and being convected downstream.
Closer to the subshock (i.e., for $z > \zLF$) the turbulence will be
intense enough to destroy any long-range correlations and Bohm-like
diffusion is assumed to occur. For $z< \zLF$, we have a mixture of
ballistic motion and diffusion, as  described below.

\subsection{Particle propagation with L\'evy-walk}\label{LFpdf}
The \mc\ method we employ numerically solves a Boltzmann equation with a collision operator which is determined by collision frequencies averaged over the self-generated background turbulence. By replacing a ``diffusion approximation" with a collision operator we are able to model \PAS\ controlled by short-scale fluctuations on the order of the particle gyroradius with arbitrary pitch angle distributions.
Therefore, particle transport is not restricted to standard diffusion
and \supDif, i.e., \levy-walk,  propagation can be directly modeled.
All that  is required for \supD\ to be accurately modeled is to define a proper particle path length probability distribution.

If path lengths at a given position are determined by macroscopic, long-range field correlations on scales larger than the
EP particle gyroradius, or by highly intermittent turbulence, then \supD\ will occur.
In the case when the path length distribution is Gaussian, or has a power-law index $\nu > 3$,  the probability of a EP having a free path that departs widely from the mean value is small and normal diffusion occurs
with a well defined {\it mean}  free path and  mean square displacement $\left < \Delta z^2 \right > = \aD t$.

On the other hand, if 2 $\leq \nu < 3$ \supDif\ propagation occurs with  $\left < \Delta z^2 \right > = \aD t^b$ where $b>1$. In this case there is a non-negligible probability for the free path to be much longer than the mean.
Physically, such a situation can be expected near the FEB where strong, unsaturated EP driven turbulence growth occurs. In this case, the turbulence is expected to be intermittent and long-range correlations are not immediately destroyed.
Closer to the subshock, \NL\ interactions of strong magnetic fluctuations are likely to smooth out the intermittency and  the downstream  turbulence is likely to be statistically homogeneous.
As a first approximation for this complicated situation, we model \levy-walk  propagation in the precursor in a region away from the subshock between $\Lfeb$ and $\zLF$. The effect of varying  $\zLF$ is examined.

\begin{figure}  % Fig 1
\includegraphics[width=3.0in]{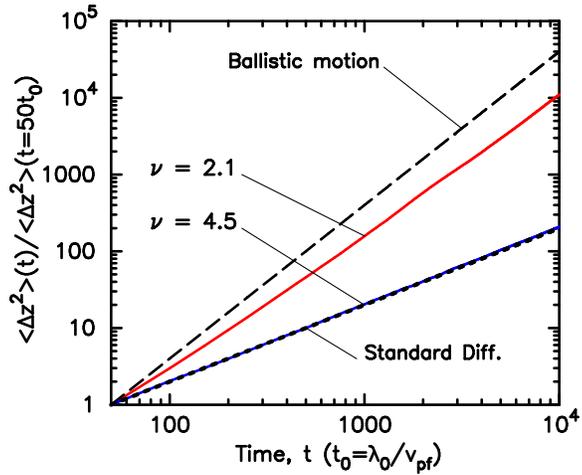}
 \caption{The mean square displacement $\left < \Delta z^2 \right >$ as a function of time for a particle which is propagating with a probability
distribution of scattering lengths as given by
Eq.~(\ref{probability_MFP}).
The displacement is the projection on one coordinate for a full 3D propagation.
The solid (red) line is the \mc\ result for $\nu$ = 2.1 which demonstrates $\left < \Delta z^2 \right > = \aD t^b$ with $b \simeq 1.76$, i.e., \supDif\ propagation.
The solid (blue) line is the result for $\nu = 4.5$. This is hardly distinguishable from the pure diffusion case (where $b = 1$) shown by a dotted curve.
The dashed (black) curve shows ballistic motion where $b = 2$.
The curves are normalized to $\left < \Delta z^2 \right >$ at $t=50 t_0$, where $t_0=\lambda_0/\vpf$ and $\vpf$ is the particle speed in the plasma frame.
These consistency checks were done using $A_{\theta}^2 = \pi^2/2$ in the \mc\ algorithm.
For the $\nu=2.1$  and 4.5 results, $10^6$ \mc\ particles of the same energy were propagated in a uniform upstream flow in order to check our \levy-walk  algorithm.
\label{fig:superdiff}}\vspace{-1.\baselineskip}
\end{figure}

Once a path length distribution is specified, as with
Eq.~(\ref{probability_MFP}), the \mc\ algorithm determines $\left < \Delta z^2 \right >(t)$
without further assumptions.
As an illustration, we show in Fig.~\ref{fig:superdiff}
\mc\ calculations with our algorithm of
$\left < \Delta z^2 \right >(t)$ vs. $t$
for two values of $\nu$, as indicated.
This calculation is done in 3D geometry and the projection onto one axis is plotted.
The \supDif\ case with $\nu = 2.1$ yields a slope $b \simeq 1.76$.
We note that in this example we restrict $\LLF$ to be equal to or greater than the Bohm limit, i.e., to the interval $[\lambda_0,\infty)$.
To test our  \mc\ algorithm we performed the simulation without this restriction, where the scattering length is allowed to populate the interval $[0,\infty)$, and we obtained
$b=1.86$ in good agreement with the scaling  $b=4-\nu =1.9$ presented by
\citep[][]{Levy_intermitt} and \citep[][]{Trotta2015}.
The solid curve labeled $\nu = 4.5$ is almost identical to the standard diffusion result shown with a dotted curve. For comparison we also  show the ballistic case with $b=2$.

Particle transport with \levy-walk is done in the following way.
When a particle is at a position upstream from
$\zLF$ a random number $\Xone$ is chosen and the particle's
scattering length, $\LLF(z,p)$, is found from
Eq.~(\ref{mfp_levy_fly_mu}).
This determines the collision time $t_c = \LLF/\vpf$.
However, since $\LLF$ depends on $\lambda_0(p,z)$ and is position dependent in the \NL\ model, its value can change during $t_c$.
To accommodate this we set the time interval
\begin{equation}
\deltime (z,p) = \frac{A_{\theta}^2 \LLF(z,p)}{6 \vpf}
\ ,
\end{equation}
during which the particle moves with a constant speed. After
$\deltime (z,p)$, the particle scatters with a new $\Lz(z,p)$. The
value of $\Xone$ is kept fixed for $\Nscat$ scatterings where
$\Nscat$ is 1 plus the integer part of $(\pi/A_{\theta})^2$. If a
particle completes $\Nscat$ scatterings without escaping, the cycle
is repeated with a new $\Xone$. In the simulations presented below
we set the parameter $A_{\theta}^2 = \pi^2/2$. This large-angle scattering value was justified in
\citep{EBJ96}.

\section{Results} \label{sec:results}
In most of the following examples we calculate \MFA\ with \supDif\ EP
propagation in \NL\ shocks where the energy and momentum conserving
shock structure has been determined \SCly. However, in order to
isolate and highlight the effects of \supD, we discuss some
unmodified {(UM)} shocks with a discontinuous bulk-flow-velocity
profile where energy and momentum are not conserved. For these unmodified 
shocks, $\LLF(z,p) =\Lz(p)$, i.e., spatially independent Bohm diffusion.
These unmodified solutions are not, of course, intended to represent physical models.

In all cases, the shock speed $u_0=5000$\,\kmps, the far upstream plasma density $n_0=0.3$\,\pcc, the background magnetic field $B_0=3$\,\muG, and we accelerate only protons. The \Facc\ is limited by an upstream FEB at $\Lfeb=-10^8\,\rgz$, where
$\rgz= m_p u_0c/(eB_0) \simeq 5.6 \times 10^{-9}$ pc.

As described in \S~\ref{sec:SD}, EPs move super-diffusively in
the shock precursor between $\Lfeb$ and $\zLF$ with a scattering
length $\LLF(z,p)$ given by the \levy-walk probability distribution
Eq.~(\ref{mfp_levy_fly_mu}) with $\nu =2.1$.
The specific value $\nu=2.1$ is chosen because it is close to the Cauchy distribution and  represents all of the features typical for prominent \supDif\ propagation while having a finite scattering length.

We show examples with $\zLF=-10^4$, $-10^5$, and $-10^6\,\rgz$.
For $z>\zLF$, including downstream from the subshock, the EP scattering length is diffusive, i.e., $\LLF(z,p) = \Lz(z,p)$. For all \NL\ calculations,  $\Lz(z,p)$ is determined with MFA from Bell's instability, the
resonant streaming instability, the non-resonant long-wavelength instability, and the mirror instability described here for the first time.
For the parameters used here, the differences in $\Lz(z,p)$ derived with the additional mirror instability are modest except at the highest particle energies.

\begin{figure}  % Fig 2
\includegraphics[width=3.0in]{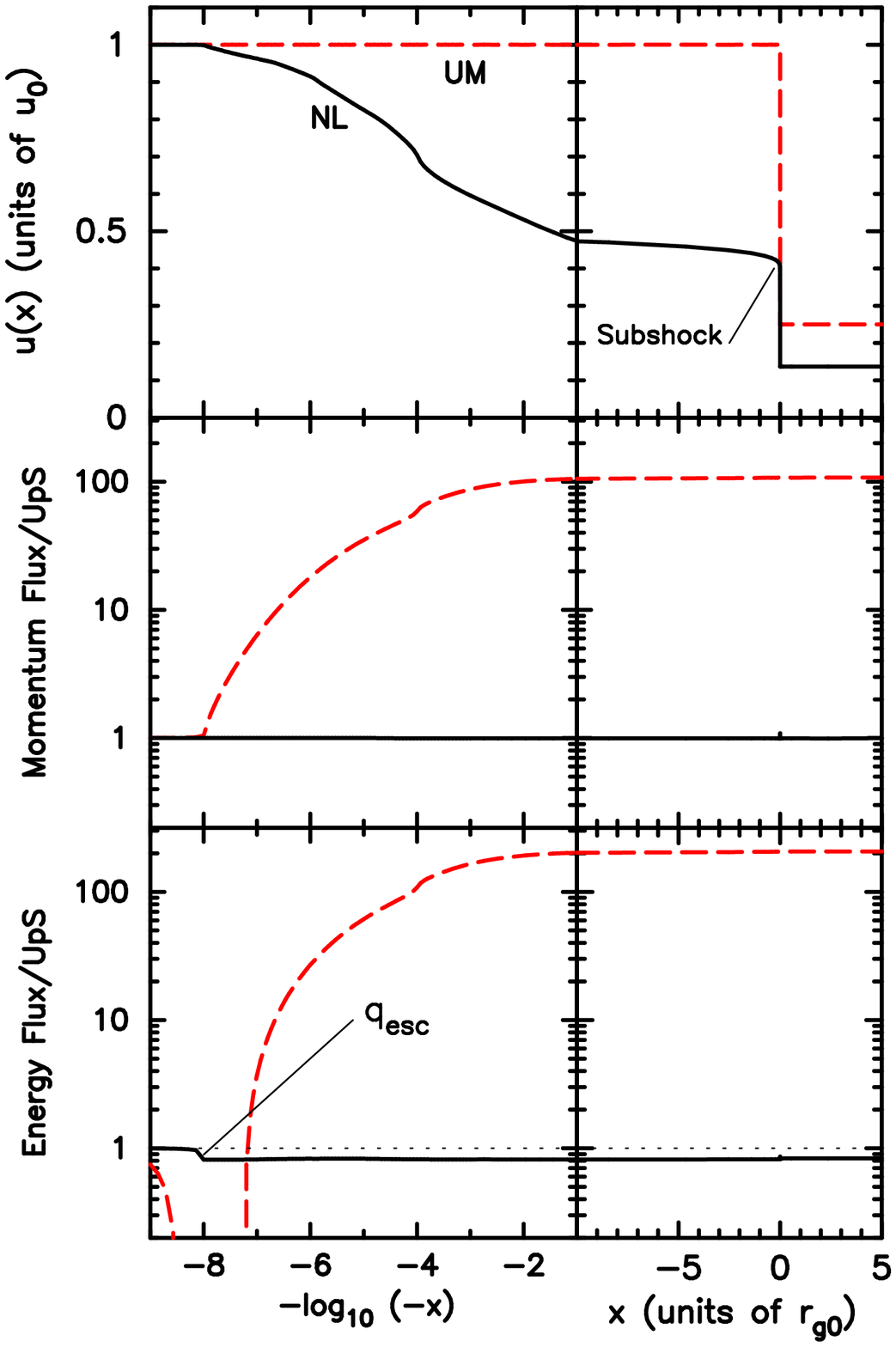}
 \caption{In all panels the dashed (red) curves show the results for an unmodified
shock with $\Rtot=4$. The solid (black) curves show \SC\ results
where the momentum and energy fluxes are conserved across the shock including the total escaping energy flux $\qesc$, i.e., $\Qesc(p)$ summed over $p$.
For this example, where all four instabilities are active, the \SC\
compression ratio is $\Rtot \simeq 7.2$ and $\sim 20$\% of the energy
flux is lost at the FEB at $z=-10^8\,\rgZ \sim -0.56$\,pc.
The subshock with $\Rsub \sim 3$ is indicated in the upper right-hand panel. All quantities are scaled to far upstream values and note the split log--linear
$x$-axis. \label{fig:conservation}} %%\vspace{-1.\baselineskip}
\end{figure}

\subsection{Nonlinear shock structure}
In the top panels of  Fig.~\ref{fig:conservation} we show the \NL\ shock structure (solid black curve) in terms of the bulk plasma flow speed $u(x)$. The solid black curves in the middle and bottom panels show the momentum and energy fluxes for the \SC\ shock. For illustration, the dashed red curves show the corresponding quantities for an unmodified shock.
An important element of the \mc\ simulation is that it contains a direct model of thermal leakage injection. The scattering assumptions described in
\S~\ref{sec:model} determine the fraction of shock-heated particles that are injected into the Fermi acceleration mechanism. This, in turn, influences the overall acceleration efficiency in an internally \SC\ fashion.

As is clear from the dashed curves in Fig.~\ref{fig:conservation}, the \mc\ injection model
predicts efficiencies that do not conserve energy and momentum in unmodified shocks. A consistent solution can be found without modifying the injection model by modifying the shock structure, as shown with the solid black curves in Fig.~\ref{fig:conservation}. As mentioned in 
\S~\ref{sec:III}, the \NL\ bulk flow speed $u(z)$ is determined by iteration and results in momentum and energy conservation to within a few percent.

In this case, momentum and energy are conserved while still having a
large  \Facc\ efficiency. We emphasize
that regardless of the injection process, shock modification must
occur if \Facc\ is  efficient. Furthermore, there must be a
corresponding increase in the overall shock compression ratio,
$\Rtot$, above the \RH\ value of $\Rtot \simeq 4$ for high Mach
number shocks. The compression  ratio is determined by the ratio of
specific heats and the energy flux leaving the shock at the FEB. The distribution of escaping EPs is
\begin{equation}
\Qesc(p) = - \frac{J(\Lfeb,p)p^4}{4\pi m_p c u_0},
\label{Qesc}\end{equation}
where $J(\Lfeb,p)$ is the EP current at $\Lfeb$ measured in
the upstream rest frame.
Using the full anisotropy information provided by the \mc\ model,
we define
the position and momentum dependent EP current as
\begin{equation}
J(z,p) = 2\pi \int_{-1}^{1}d\mu\, v \mu \FpfT \ , \label{Jzp}
\end{equation}
where $\FpfT$ is the distribution function of accelerated particles,
per $d\mu$,
 in the local rest frame  of the background plasma,
$\mu=\cos\theta$, and $\theta$ is the angle between a particle's
momentum and the $z$-axis.
The bottom panels in Fig.~\ref{fig:conservation} show that $\sim
20\%$ of the far upstream energy flux
$\qesc = \int J(\Lfeb,p)E(p) p^2dp$
(where $E(p)$ is the particle energy) escapes at $z=\Lfeb$ and the
plot for $u(x)$ shows $\Rtot \simeq 7.2$.
While $\Rtot$ increases above the \TP\ \RH\ value, the subshock
(indicated in the top  right-hand panel) must decrease 
below the  \TP\
value. For the \NL\ shock in Fig.~\ref{fig:conservation}, $\Rsub
\simeq 3$. These modifications to the shock structure from efficient
\DSA\ produce the non-power-law behavior in the \NL\ distribution
functions we discuss next.

\begin{figure}  % Fig 3
\includegraphics[width=3.25in]{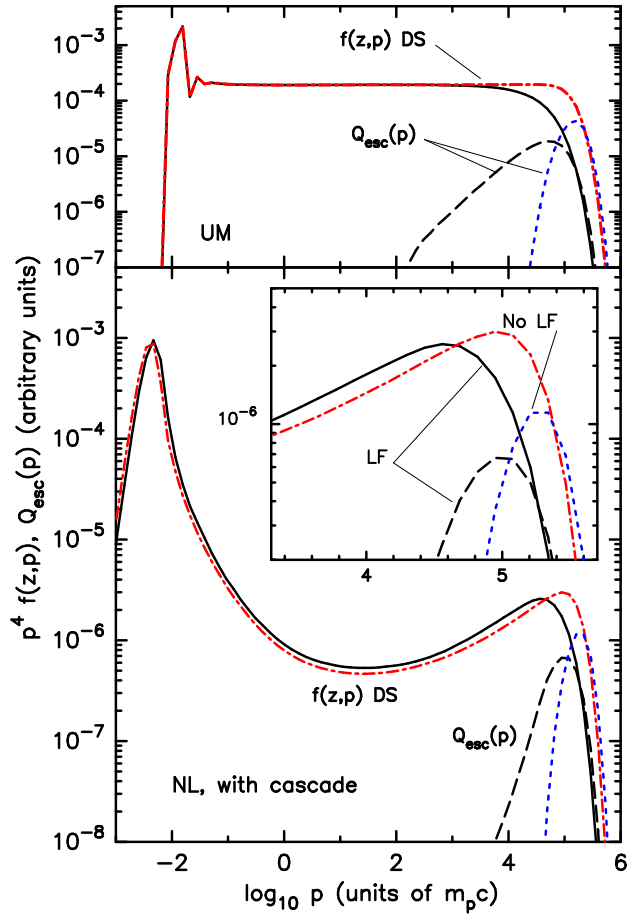} %%%{fp_UM_NL_A.eps}
\caption{Shown are proton phase-space distributions measured  in the shock
rest frame. Downstream spectra are plotted, as are distributions of particles escaping at the upstream FEB (i.e., Eq.~\ref{Qesc}), as indicated.
Spectra for unmodified (UM) shocks are in the top
panel while those for consistent \NL\ (NL) shocks are in the bottom panel
and insert. All spectra are absolutely normalized relative to each
other.
In the \levy-flight examples (solid and dashed black curves),
\supDif\ EP propagation occurs between the FEB at $\Lfeb =-10^8\rgz \sim 0.56$\,pc
and
$\zLF =-10^4\rgz$.
The normal diffusion cases are shown with dot-dashed red curves and dotted blue curves. We have included turbulence cascade in the \NL\ cases. For the unmodified shocks, the scattering is uniform without magnetic field growth.
\label{fig:fp_UM_NL}} %%\vspace{-1.\baselineskip}
\end{figure}

\begin{figure}  % Fig 4
\includegraphics[width=3.25in]{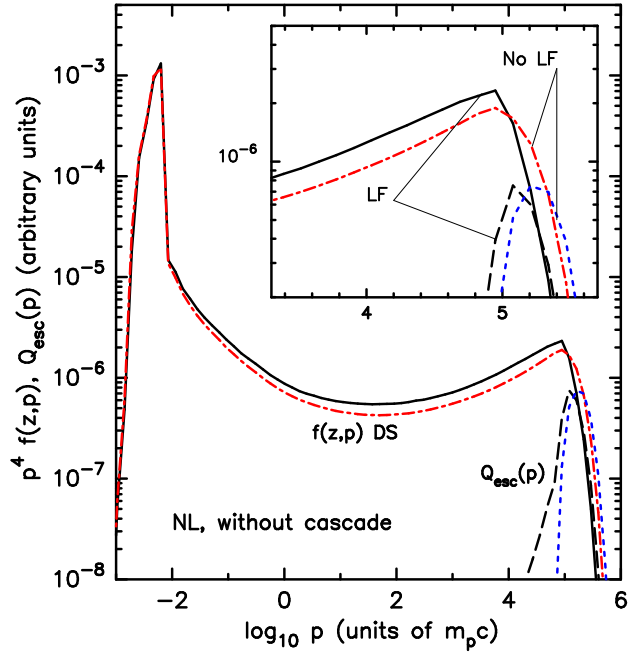} %%{fp_No_cas.eps}
\caption{Same as bottom panel in Fig.~\ref{fig:fp_UM_NL} without turbulence cascade.
In the \NL\ (NL) \levy-flight examples (solid and dashed black curves),
\supDif\ EP propagation occurs between the FEB at $\Lfeb =-10^8\rgz \sim 0.56$\,pc
and
$\zLF =-10^4\rgz$.
The normal diffusion cases are shown with dot-dashed red curves and dotted blue curves.
\label{fig:fp_NoCas}} %%%\vspace{-1.\baselineskip}
\end{figure}

\subsection{Particle spectra} \label{sec:spec}
In Fig.~\ref{fig:fp_UM_NL} we show particle spectra measured
downstream (DS) from the subshock and at  the upstream FEB as
indicated. The top panel shows spectra  for unmodified shocks while spectra in the bottom
panel are for \SC, \NL\ shocks, both with  the \supD\ parameter
$\nu=2.1$ in Eq.~(\ref{mfp_levy_fly_mu}).
The proton distributions are calculated for $\zLF= -10^4\,\rgz$ (black, solid and dashed curves) and without \levy-flight transport (red, dot-dashed and blue, dotted curves).
\SupD\ is eliminated by placing $\zLF$ outside of the FEB.
In the unmodified examples, $\Lz$ was assumed to be spatially independent
Bohm diffusion.
The examples in Fig.~\ref{fig:fp_UM_NL} are calculated with cascading while those in Fig.~\ref{fig:fp_NoCas} are calculated 
without cascading.
We note that the statistical errors in 
Figs.~\ref{fig:fp_UM_NL} and \ref{fig:fp_NoCas} are small. Except for the escaping particles,  variations are typically less than the line thickness.

These examples show that \supD\ produces a high-energy  cutoff in
$f(z,p)$ that  is broader than that for the diffusive case and
occurs at a lower momentum. This is clearly reflected in the
escaping distributions where the black dashed curves are with \supD\
and the dotted blue curves are without.
The broad cutoff  results as particles with long \levy-walk   scattering
lengths leave the FEB from deep within the precursor.
It is significant that the broadening, while still present, is less in the \SC\ shocks.
While not shown for clarity, we find that the cutoff is only weakly dependent on
$\zLF$.

\begin{figure}  % Fig 5
\includegraphics[width=3.0in]{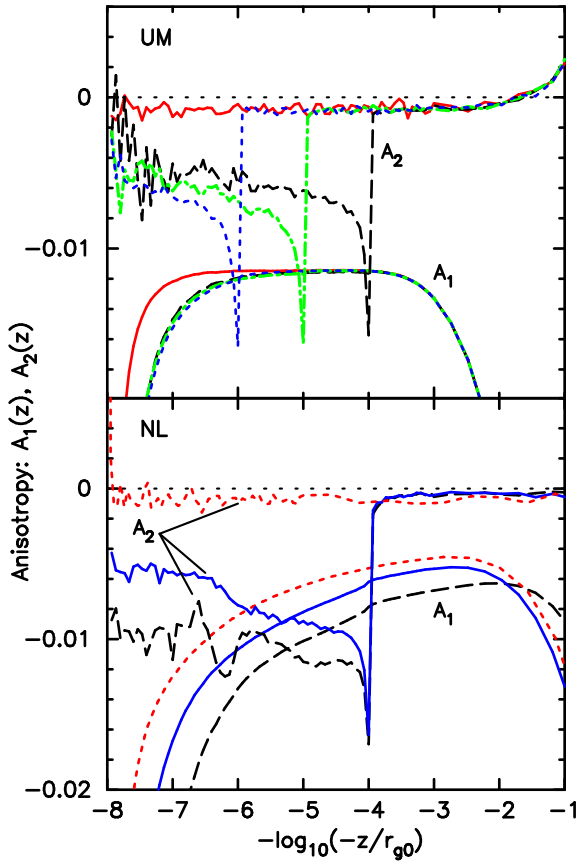}
\caption{Proton anisotropies,
as defined in Eqs.~(\ref{A1}) and (\ref{A2f}),
in the rest
frame of the background plasma.
The top panel, for unmodified (UM) shocks, shows examples where  \supD\ with $\nu=2.1$ occurs between the FEB and the upstream position $\zLF = -10^4\,\rgz$ (black dashed curves), $\zLF = -10^5\,\rgz$ (green dot-dashed curves),
and $\zLF = -10^6\,\rgz$ (blue dotted curves).
The bottom panel shows the anisotropy for \NL\
shocks with cascading (blue solid curves) and without (black dashed curves) for $\zLF=-10^4\,\rgz$.
Results with no \supD\ are shown   with cascading (red dotted curves).
The fluctuations in the $A_2$ results at large $|z|$ are  statistical errors from the \mc\ simulation.
\label{fig:anis}} %%%\vspace{-1.\baselineskip}
\end{figure}

\subsection{Particle anisotropy with super-diffusion} \label{sec:anis}
The \EP\ current anisotropy, $A_1(z)$, is given by
\begin{equation} \label{A1}
A_1(z) = N^{-1} \int_{0}^{\infty}J(z,p)v^{-1}p^2dp \ .
\end{equation}
The partial anisotropy, $A_2(z,p)$, associated with the second
spherical harmonic of  the particle distribution is defined as
\begin{equation}\label{A2_x_p}
A_2(z,p) =
\pi \int_{-1}^{1}(3\mu^{2}-1) \FpfT\, d\mu
\ ,
\end{equation}
and integrating over momentum gives
\begin{equation} \label{A2f}
A_2(z)= N^{-1} \int_{0}^{\infty}A_2(z,p)p^2dp
\ ,
\end{equation}
where
\begin{equation} \label{Nzp}
N = 2\pi \int_{0}^{\infty} p^2dp \int_{-1}^{1} d\mu\, \FpfT \ .
\end{equation}
The anisotropies are both defined in the local plasma frame
at all $z$.

In Fig.~\ref{fig:anis} we show dimensionless $A_1(z)$ and
$A_2(z)$ for the unmodified  shocks (top
panel) and \NL\ shocks (bottom panel) discussed in
Fig.~\ref{fig:fp_UM_NL}. In the unmodified case, the bulk velocity profile
is fixed with $\Rtot=4$ (see the dashed curves in
Fig.~\ref{fig:conservation}) and the  magnetic fluctuations spectra
are also fixed, are position independent, and assume a Bohm-type
scattering length $\Lz \propto p$.
The \NL\ shocks are fully consistent in shock structure, $\Rtot$,
self-generated magnetic turbulence, scattering length determination,
and scattering center speed relative to the  bulk speed of the background plasma (see \S~\ref{sec:ScatV}).

The important result here is that super-diffusive propagation
with a \levy-type distribution of particle scattering lengths results in  second harmonics much stronger than produced with diffusive propagation.
The magnitude of $A_2(z)$ is within a factor of a few to that of $A_1(z)$ in regions where super-diffusion is acting, while it is orders of magnitude less in regions with only diffusive propagation.

The second
harmonic $A_2(z)$ is negative showing that the magnetic field partial pressure transverse to the local mean field is greater than the parallel partial pressure.
This EP anisotropy will drive the so-called mirror
instability, as we describe in more detail in Appendix~\ref{sec:A1}.
The mirror instability is non-resonant where the
growing magnetic fluctuations are nearly transverse wavevectors
$2k_{\perp}^{2}>k_{\parallel}^{2}$ of scales larger than the
gyroradius of particles, $\Rgp$, which are contributing into the
transverse pressure,
i.e., $k_{\perp} \Rgp <1$ but $k_{\perp}\lambda(p) >1$.
The mirror mode is compressive and long-wavelength in the sense that $k_{\perp} \Rgp <1$. This characteristic may help increase the efficiency of scattering at the highest energy end of the accelerated particle
spectrum.

We note that even though we emphasize effects produced for extremely efficient \DSA\ producing hard, concave spectra, as shown in
Figs.~\ref{fig:fp_UM_NL} and \ref{fig:fp_NoCas}, the growth rates we derive can be applied to weaker shocks. The mirror instability growth rate formulation can be applied to steep spectra and non-power-laws.

\subsection{Magnetic field amplification} \label{sec:mfa}
The \EP\ current and the quadrupole anisotropy, demonstrated in
Fig.~\ref{fig:anis}, drive resonant, short--, and long--wavelength
instabilities, as well as the mirror instability first discussed
here.
This MFA is included in our \NL\  model.
The turbulence growth rates and the transport equations used were
discussed in detail in sections 2.4 and 2.5 of \citep{beov14}, while the growth rate for the mirror
instability which is associated with the quadrupole anisotropy of
EPs produced by super-diffusive propagation is presented in
Appendix~\ref{sec:A1}.

The strong \NL\ aspects of MFA force a \SC\ description of the energy exchange between the EPs and the  magnetic field, as well as with the bulk shock flow.
We include the cascade of turbulence energy to shorter wavelengths
taking into account the energy
dissipation and heating of the background plasma. For comparison, we include models without turbulence cascade and show that cascading
influences the magnetic fluctuation spectra,
the total shock compression, and the downstream proton temperature.

\begin{figure}  % Fig 6
\includegraphics[width=2.8in]{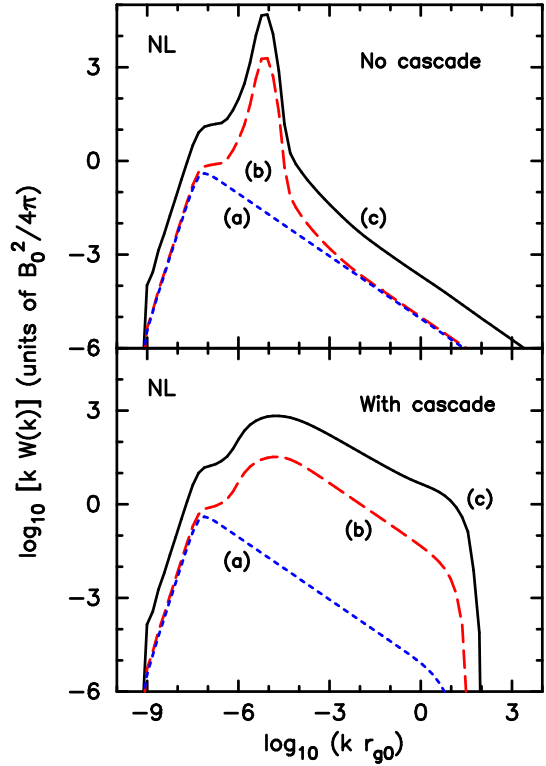}
\caption{Spectral energy densities of the EP-driven magnetic
fluctuations measured at three positions: (a) the FEB, (b) $z = 0.01
\Lfeb$, and (c) in the downstream region. There is a strong effect
on $kW(k)$ from turbulence cascade. For both of these \NL\ shocks,
$\zLF=-10^4\,\rgz$ and $\nu=2.1$.
\label{fig:turb}} %%%\vspace{-1.\baselineskip}
\end{figure}

In Fig.~\ref{fig:turb} we show the self-generated magnetic turbulence spectra, with and  without cascading, at various positions relative to the subshock at $z=0$. These models include \supD\  beginning at $\zLF= -10^4\,\rgz$.
The top panel, without cascade, shows a strong spike in wave power at long wavelengths resulting from \supD. With cascade, this turbulent energy is effectively shifted to shorter wavelengths.

\begin{figure}  % Fig 7
\includegraphics[width=2.9in]{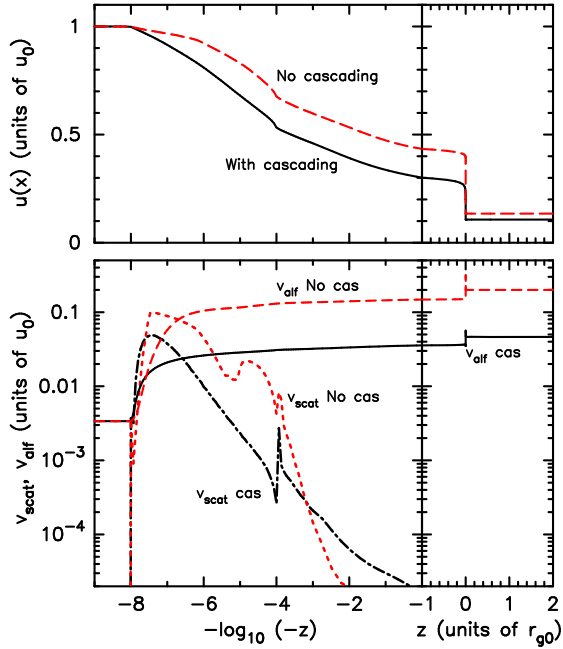}
\caption{The bulk flow speed (top panel) and scattering center velocity derived from the
\NL\ \mc\ model (bottom panel) with and without turbulent cascade, as indicated. Also shown in the bottom panel is $\valf(\Bls)$,
the local \alf\ speed derived using the amplified magnetic field
(dashed curves). For these examples, $\zLF=-10^4\,\rgz$.
\label{fig:vscat}} %%%\vspace{-1.\baselineskip}
\end{figure}

\subsection{Effective Scattering Center Velocity} \label{sec:ScatV}
The magnetic fluctuations produced by EP-driven instabilities in the
shock precursor move relative to the bulk plasma with a speed $\vscat(z)$. This is a highly \NL\ effect since $\vscat(z)$ directly influences the effective compression ratio for \DSA\ which, in turn, determines the \Facc\ efficiency and the MFA.
While virtually all work on \DSA\ attempting to consider a finite $\vscat(z)$ assume it is some function of the \alf\ speed, $\valf$, there is no justification for this assumption other than ease of computation.
A linear analysis of the four instabilities included in our model
shows a vast variety of phase velocities. Significantly, even though
the fastest growing EP-driven modes are highly anisotropic, the
non-resonant Bell mode, as well as the long-wavelength and mirror
modes, have phase speeds that are typically well below the local
\alf\ speed calculated with the local amplified large-scale magnetic
field \citep[see Fig.~18 in][]{beov14}.

The flexibility of the \MC\ method allows us to determine $\vscat(z)$ from macroscopic energy conservation without making any assumptions regarding the amplified field.
In Fig.~\ref{fig:vscat} (bottom panels) we show $\vscat(z)$ for the \NL\
cascade and non-cascade cases shown in Fig.~\ref{fig:turb}.

As is clear from Fig.~\ref{fig:vscat}, $\vscat$ is different in magnitude and spatial structure from $\valf$. Near the FEB $\vscat > \valf$ but closer to the subshock $\vscat$ can be orders of magnitude smaller.
It is important to
note that even though $\Vscat(z)$ may be small relative to both the
bulk  flow speed, it has a strong effect on
the energy exchange between the accelerated particles, the bulk shock flow, and the magnetic field amplification.
As in \citep{beov14}, $\vscat(z)$ is determined consistently with the
shock structure modified by energetic particles and the magnetic field amplification.

\section{Summary and conclusions}
In collisionless astrophysical plasmas,  particle transport is
determined by charged particles interacting with magnetic turbulence
and coherent magnetic structures over a broad wavelength range.
These interactions are essential for collisionless shocks to form
and accelerate particles to high energies. Details of
the wave-particle interactions will influence energetic particles observed at Earth as
well as radiative signatures of specific objects.
While normal diffusion and advection have been proven to determine
the long-range   particle transport in quasi-homogeneous magnetic
turbulence, studies both in the laboratory and in space plasmas have
revealed a variety of sub- and super-diffusive regimes where
particle transport may significantly depart from standard diffusive
propagation for intermittent turbulence (e.g., during the growth of long-wavelength magnetic fluctuations in
shock precursors).

We have presented the first \NL\ calculation of efficient  Fermi
shock acceleration that includes \supD\ in a consistent manner. The
\mc\ simulation we use, since it does not make a diffusion
approximation, can model highly anisotropic particle distributions
and magnetic field amplification and is well suited for these calculations.
It includes  \NL\ effects from thermal particle injection, shock
modification, the self-generation of magnetic turbulence, turbulence
cascade, and a consistently determined scattering center speed.
While we fully expect that future
large-scale PIC simulations will necessitate a refinement of our assumptions, accounting for important  multi-dimensional effects is well beyond current PIC capabilities.

We show that \supDif\ particle transport in the shock precursor produces
specific, anisotropic EP distributions which are
characterized by a pronounced quadrupole anisotropy where the
transverse particle pressure dominates the parallel particle
pressure. We show that this type of anisotropy results in a mirror-like
instability which is most prominent when the shock accelerated spectrum
is $f(p) \propto p^{-4}$ or harder near the maximum energy of the
accelerated particles (before the spectral break), as is  expected
for high Mach number shocks.
Magnetic field and plasma
structures produced by the mirror instability in cosmic plasmas were
observed in planetary magnetosheaths
(e.g., \cite{2011PhRvL.107x5005H}) and may appear as well in high
resolution {\sl Chandra} images of young supernova remnants where
synchrotron structures tracing the magnetic fields are prominent
\cite{2014RPPh...77f6902T}.

\SupD\ also results in a  broadening of the spectrum of EPs that
escape from the shock precursor since there is a significant
probability that EPs can leave the shock from deep inside the
precursor. This broadening may impact models of \gamray\ production
by shocks interacting with dense molecular clouds and  modify
predictions for the integrated spectra of energetic particles.
\\
\\
\noindent {\bf ACKNOWLEDGMENTS} A.M.Bykov and S.M.Osipov acknowledge
support from RSF grant 16-12-10225. D.C.Ellison acknowledges support
from NASA grant NNX11AE03G. Discussions concerning this work were
held at the Aspen Center for Physics which is supported by National
Science Foundation grant PHY-1066293.

\bibliographystyle{aa}
%\bibliography{\bibliographystyle{mn2e}}
%\bibliography{bib_DCE}

%% bbbb  Note: must have files:  aa.bst  and  aa.cls
%% \bibliographystyle{aa} % A&A style
%%\bibliography{c:/a_a_TOP/bibTeX/bib_DCE}
\bibliography{biblio_dce_pre}

\appendix

\section{Mirror instability of anisotropic EP distributions} \label{sec:A1}
As mentioned in \S~\ref{sec:anis}, the \mc\ method follows an
arbitrary angular distribution of accelerated particles and thus can
determine the second spherical harmonic, $A_2(z,p)$, of the  EP
distribution at all positions across the \NL\ shock structure.
A non-zero $A_2(z,p)$  indicates that there is a pressure
anisotropy and this may result in the growth of a firehose or
mirror instability depending on the ratio of partial pressures along
and transverse to the mean magnetic field
\citep[e.g.,][]{Treumann97}.

It has long been suggested that the mirror instability occurs in Maxwellian plasmas when the
transverse pressure dominates the parallel pressure, as
discussed by \citep[][]{CKW58_mirror,sagdeev_mirror61,1969PhFl...12.2642H,
SouthwoodKivelson93,2011PhRvL.107x5005H},
and many others.
Observational signatures of the mirror
instability have been found in cosmic
plasmas \citep[e.g.,][]{mirror_TS_11} as well.
However, in highly nonthermal systems like the collisionless shocks discussed in this paper the non-thermal relativistic particle pressure  is important. In the downstream flow it is comparable to the thermal pressure and it can be well above the thermal particle pressure in the shock precursor.

Therefore, consistent \NL\ shock solutions must determine the growth rate of the mirror instability, in the thermal background plasma, driven by a superthermal particle pressure anisotropy that occurs with \supDif\  transport.
Our derivation treats the accelerated particles, the source of free
energy for the mirror instability, kinetically, while the background
plasma is described with an  MHD approximation
\citep[e.g.,][]{bbmo13}.
Since injection is a continuous process in the \mc\ simulation, all
particles that make at least one crossing of the subshock (at $z=0$)
from downstream to upstream are included as superthermal.

For the background plasma we have
\begin{eqnarray} \label{eqMotiontot}
& & \rho\left(\frac{\partial\mathbf{u}}{\partial
 t}+(\mathbf{u}\nabla)\mathbf{u}\right)
 =- \nabla p_{g}+  \frac{1}{4\pi}(\nabla\times\mathbf{B})\times\mathbf{B}-
 \nonumber \\
& & - \frac{1}{c}(\jCR -e \nCR \mathbf{u})\times\mathbf{B}-\int\mathbf{p}I [\FcrUp]d^{3}p
\ ,
\end{eqnarray}
where $\rho$, $\mathbf{u}$, and $p_{g}$ are  the background plasma density,
macroscopic velocity, and pressure respectively.
In addition, the continuity equation is
\begin{equation}\label{EqDensCont}
\frac{\partial\rho}{\partial t}+\nabla (\rho\mathbf{u} ) = 0
\ ,
\end{equation}
the  electric and magnetic fields  $\mathbf{E}$ and $\mathbf{B}$
satisfy
\begin{equation}\label{largeInd}
\frac{\partial\mathbf{B}}{\partial t}=\nabla \times(\mathbf{u}\times
\mathbf{B})
\ ,
\end{equation}
and
\begin{equation}\label{KinEqCR1}
\frac{\partial \FcrUp}{\partial t}+\mathbf{v}\cdot\frac{\partial
\FcrUp}{\partial\mathbf{r}}+e\mathbf{E}\cdot\frac{\partial
\FcrUp}{\partial\mathbf{p}}-\frac{ec}{\mathcal{E}}\mathbf{B}
\cdot\widehat{\mathbf{\mathcal{O}}}\FcrUp  =I[\FcrUp]
\ ,
\end{equation}
where $\FcrUp$, $\nCR$, and $\jCR$  are the EP distribution
function, EP concentration, and electric current, $\mathcal{E}$ is
the EP particle energy,
$\displaystyle\widehat{\mathbf{\mathcal{O}}}=\mathbf{p}\times\frac{\partial
}{\partial \mathbf{p}}$ is the momentum rotation operator, $c$ is
the speed of light, $e$ is the particle  charge, and
$\displaystyle\Omega = eB_{0}c/\mathcal{E}$ is the EP particle
gyro-frequency.

The right-hand-side of Eq.~(\ref{KinEqCR1}), $I[\FcrUp]$,  is
the collision operator
describing EP interactions with magnetic fluctuations carried by the
background plasma.
In the background plasma frame,
\begin{equation}
I[\FcrUp] = -\nu_c\left(\FcrUp - \FcrIso \right) \ ,
\end{equation}
where $\FcrIso$ is the isotropic part of the distribution function,
$\nu_c = \epsilon \Omega$  is the EP scattering frequency  by
magnetic fluctuations with $\epsilon \leq 1$, and $\int\mathbf{p}
I[\FcrUp] d^3p = - \epsilon B_0 \jCR/c$ in Eq.~(\ref{eqMotiontot}).

In the rest frame of the background plasma we impose small
perturbations of the local plasma parameters in
Eq.~(\ref{eqMotiontot})--(\ref{KinEqCR1}) as
$\xi=\xi_{0}+\delta\xi$,
with
$\delta\xi\propto \exp(i\mathbf{k}\mathbf{r}-i\omega t)$.
Assuming the adiabatic equation for the background plasma,
$\nabla \delta p_{g}=a_{0}^{2}\nabla \delta\rho$,
with
$a_0 = \sqrt{\gamma_g p_0/\rho_0}$,
where $\gamma_{g}$ is the adiabatic index.
Furthermore, we define $\mathbf{e}_{z} = \mathbf{B}_0/B_0$ as the
parallel direction, $\mathbf{e}_x$ is the transverse  direction, and
$\mathbf{k} = k_\parallel \mathbf{e}_z + k_\perp \mathbf{e}_x$.

Since we are only  concerned here with instabilities due to the
anisotropic EP pressure (i.e., the quadrupole anisotropy of the EP
distribution), we consider the unperturbed EP distribution function
with no mean electric current \citep[see][ for discussions of the
resonant and current driven
instabilities]{schureea12,bbmo13,marcowithea16}.
With no mean current, the quadrupole anisotropy has the form
\begin{equation}\label{distrF0}
f^{ep}_{0}\left(p,\mu\right)=\frac{\nCR N\left(p\right)}{4\pi}\left[1+\frac{\chi}{2}\left(3\mu^{2}-1\right)\right],
\end{equation}
where
$\int_0^\infty N(p) p^2 dp = 1$ and $\chi$ is a
quadrupole anisotropy parameter with $|\chi|<1$.

Keeping only linear responses to the perturbations in
Eq.~(\ref{eqMotiontot})--(\ref{EqDensCont}), we obtain
\begin{eqnarray} \label{Eqbx}
& & \left(\omega^{4}-\omega^{2}\left(v_{a}^{2}+a_{0}^{2}\right)k^{2}+v_{a}^{2}a_{0}^{2}k^{2}k_{\parallel}^{2}\right)\delta B_{x} \nonumber \\
& &
=\left(\omega^{2}-a_{0}^{2}k_{\parallel}^{2}\right)i\frac{B_{0}^{2}k_{\parallel}}{c\rho_{0}}\left(\delta
j^{ep}_{y}-\epsilon\delta j^{ep}_{x}\right)  \nonumber \\
& &
-ia_{0}^{2}k_{\parallel}^{2}\frac{B_{0}^{2}k_{\perp}}{c\rho_{0}}\epsilon\delta
j^{ep}_{z},
\end{eqnarray}
and
\begin{equation}\label{Eqby}
\left(\omega^{2}-v_{a}^{2}k_{\parallel}^{2}\right)\delta
B_{y}=-i\frac{B_{0}^{2}k_{\parallel}}{c\rho_{0}}\left(\delta
j^{ep}_{x}+\epsilon\delta j^{ep}_{y}\right),
\end{equation}
where $v_a = B_0/\sqrt{4 \pi \rho_0}$ is
the \alf\ velocity.
The linearized Eq.~(\ref{KinEqCR1}) has the form
\begin{eqnarray} \label{KinEqCRLin}
& &
\left[\nu_c+i\left(-\omega+k_{\parallel}v\cos\theta\right)\right]\delta
f^{ep}+ik_{\perp}v\sin\theta\cos\varphi\delta f^{ep}-
 \nonumber \\
& & -\Omega\frac{\partial \delta f^{ep}}{\partial
\varphi}=-e\delta\mathbf{E}\cdot\frac{\partial
f^{ep}_{0}}{\partial\mathbf{p}}+\frac{ec}{\mathcal{E}}\delta\mathbf{B}\cdot\widehat{\mathbf{\mathcal{O}}}f^{ep}_{0},
\end{eqnarray}
where $\theta$ and $\varphi$ are the pitch and azimuthal angles
between the EP particle velocity and the direction $\mathbf{e}_{z}$,
correspondingly.

We first consider the weakly collisional case with $\epsilon \ll 1$.
In the long-wavelength regime, $k_\parallel v /\Omega \ll 1$,
$k_\perp v /\Omega \ll 1$, and $\omega/\Omega \ll 1$.
Then one obtains the response of the superthermal particle current  $\delta \mathbf{j^{ep}}$ on the magnetic field perturbation
$\delta\mathbf{B}$ in the form
\begin{eqnarray} \label{JxLongWave}
& & \delta j^{ep}_{x}=\frac{\pi e}{2B_{0}}\int dp
p^{2}\int_{-1}^{1}d\mu v\left(1-\mu^{2}\right)\frac{\partial
f_{0}^{ep}\left(p,\mu\right)}{\partial \mu}
 \nonumber \\
& & \left\{2\delta B_{x}-2i\frac{k_{\parallel}v\mu}{\Omega}\delta
B_{y}\right\},
\end{eqnarray}
and
\begin{eqnarray} \label{JyLongWave}
& & \delta j^{ep}_{y}=\frac{\pi e}{2B_{0}}\int dp
p^{2}\int_{-1}^{1}d\mu v\left(1-\mu^{2}\right)\frac{\partial
f_{0}^{ep}\left(p,\mu\right)}{\partial \mu}
 \nonumber \\
& & \left\{2\delta B_{y}+2i\frac{k_{\parallel}v\mu}{\Omega}\delta
B_{x}-i\frac{k_{\perp}^{2}v^{2}\left(1-\mu^{2}\right)}{\Omega\left(k_{\parallel}v\mu-\omega\right)}\delta
B_{x}\right\}.
\end{eqnarray}

Consider the brace in Eq.~(\ref{JyLongWave}). With the assumption for the unperturbed quadrupole
anisotropy given by Eq.~(\ref{distrF0}), the second and third terms  in the brace are of the same order of magnitude. The second term gives the well known firehose instability which grows if $\chi>0$, while the third term results in the mirror instability if $\chi<0$.
Using Eqs.~(\ref{distrF0}) and (\ref{A2_x_p}), the parameter $\chi$ can be connected to $A_2(p)$ with
$A_2(p) = \nCR N(p) \chi/5$.
Note that if the EP scattering rate by magnetic turbulence $\nu_c
\sim \Omega$, which may occur if $\epsilon \sim$ 1, then the mirror
instability is suppressed.

Integrating  Eq.~(\ref{JyLongWave}) over $\mu$, with account taken of
Eq.~(\ref{distrF0}), one obtains
\begin{eqnarray} \label{Amir}
& & \Amir (\tau)=\int_{-1}^{1}\frac{\left(1-\mu^{2}\right)^{2}\mu
d\mu}{1-\tau\mu}=-\frac{16}{15\tau}+\frac{10}{3\tau^{3}}-\frac{2}{\tau^{5}}+
 \nonumber \\
& & +\frac{1}{\tau^{2}}\left(1-\frac{1}{\tau^{2}}\right)^{2}\ln
\left|\frac{\tau+1}{\tau-1}\right|-
 \nonumber \\
& &
-i\pi\frac{1}{\tau^{2}}\left(1-\frac{1}{\tau^{2}}\right)^{2}\Theta\left(\left|\tau\right|-1\right),
\end{eqnarray}
where $\tau = k_\parallel v/\omega$ and $\Theta(z)$ is the Heaviside step function.
The asymptotic form of Eq.~(\ref{Amir}) for $\tau\ll 1$ is
\begin{equation}\label{AmirSmallTau}
\Amir(\tau)\rightarrow \frac{16}{105}\tau+O\left(\tau^{3}\right),
\end{equation}
while, for  $\tau\gg 1$,
\begin{equation}\label{AmirLargeTau}
\Amir(\tau)\rightarrow
-\frac{16}{15\tau}+O\left(\frac{1}{\tau^{2}}\right).
\end{equation}

The response of the electric current carried by the  energetic particles in these limits is
\begin{equation}\label{JxLongWave1}
\delta j^{ep}_{x}=-i\frac{\delta B_{y}\chi
\nCR k_{\parallel}c}{5B_{0}^{2}}\int_{0}^{\infty}vp^{3}N\left(p\right)dp,
\end{equation}
and
\begin{equation}\label{JyLongWave1}
\delta j^{ep}_{y}=i\frac{\delta B_{x}\chi
n_{ep}c}{5B_{0}^{2}}\left(k_{\parallel}-\frac{2k_{\perp}^{2}}{k_{\parallel}}\right)\int_{0}^{\infty}vp^{3}N\left(p\right)dp.
\end{equation}

Let us define the unperturbed pressure
\begin{equation}\label{P0}
P_{0}=\frac{\nCR}{3}\int_{0}^{\infty}vp^{3}N\left(p\right)dp
\ ,
\end{equation}
and note that for the distribution function Eq.~(\ref{distrF0}),
\begin{equation}\label{P_paral}
P_{\parallel}=\int vp\mu^{2}
f^{ep}_{0}\left(p,\mu\right)d^{3}p=P_{0}\left(1+\frac{2\chi}{5}\right),
\end{equation}
\begin{eqnarray} \label{P_perp}
& & P_{\perp} =
\int vp \left(1-\mu^{2}\right) \cos^{2}\varphi f^{ep}_{0}\left(p,\mu\right)d^{3}p \nonumber \\
& &
=\int vp\left(1-\mu^{2}\right)\sin^{2}\varphi f^{ep}_{0}\left(p,\mu\right)d^{3}p \nonumber \\
& & =P_{0}\left(1-\frac{\chi}{5}\right),
\end{eqnarray}
and
\begin{equation}\label{delta_P_f0}
\delta P=P_{\parallel}-P_{\perp}=\frac{3\chi}{5}P_{0}
\ .
\end{equation}

Now, substituting Eqs.~(\ref{JxLongWave1}) and (\ref{JyLongWave1})
into (\ref{Eqbx}) and (\ref{Eqby}) we obtain the dispersion relations
\begin{eqnarray} \label{Eqbx1}
& & \omega^{4}-\omega^{2}\left(v_{a}^{2}+a_{0}^{2}\right)k^{2}+v_{a}^{2}a_{0}^{2}k^{2}k_{\parallel}^{2} \nonumber \\
& & =-\left(\omega^{2}-a_{0}^{2}k_{\parallel}^{2}\right)\frac{3\chi
P_{0}}{5\rho_{0}}\left(k_{\parallel}^{2}-2k_{\perp}^{2}\right),
\end{eqnarray}
and
\begin{equation}\label{Eqby1}
\omega^{2}-v_{a}^{2}k_{\parallel}^{2}=-\frac{3\chi
P_{0}}{5\rho_{0}}k_{\parallel}^{2}
\ .
\end{equation}
The dispersion equation splits into two independent equations where
 the mirror instability is determined by Eq.~(\ref{Eqbx1}) under
conditions $2k_{\perp}^{2}>k_{\parallel}^{2}$ and $\chi<0$.
For $k_{\perp}^{2}\gg k_{\parallel}^{2}$  Eq.~(\ref{Eqbx1})
simplifies to
\begin{equation}\label{Eqbx1a}
\omega^{2}-\left(v_{a}^{2}+a_{0}^{2}\right)k^{2}=\frac{3\chi
P_{0}}{5\rho_{0}}2k_{\perp}^{2},
\end{equation}
Here the perturbation of the magnetic field is mostly $\delta B_{z}$
which are connected to the background plasma density variations by
$\displaystyle\delta\rho\approx\rho_{0}\frac{\delta B_{z}}{B_{0}}$.

Then from Eq.~(\ref{delta_P_f0}), with account taken that
$k\approx k_{\perp}$,  one finally obtains
\begin{equation}\label{Eqbx1b}
\Omir^2 =\left(v_{a}^{2}+a_{0}^{2}+2\frac{\delta
P}{\rho_{0}}\right)k^{2}.
\end{equation}
Eq.~(\ref{Eqbx1b}) shows that growing modes occur 
when $\left(v_{a}^{2}+a_{0}^{2}+2\frac{\delta P}{\rho_{0}}\right)
<0$, which may happen only if $\delta P \propto \chi <0$, i.e., the
mirror instability is driven by anisotropic EP pressure.

\end{document}